\documentstyle[aps]{revtex}
\begin{document}
\draft
\title{Quantum diffusion at high interstitial concentrations: Application to surface diffusion}
\author{G.L Buchbinder and V.N. Shapov}
\address{Department of Physics, Omsk State University, Peace Avenue 55-a 644077, Omsk, Russia}
\date{\today}
\maketitle
\begin{abstract}
In this work we first study the quantum  diffusion in a volume of a crystalline
solid at  high interstitial concentrations when  the effects of the short-range
interactions between the diffusing particles are to be the factors. Within the scope
of the small-polaron formalism the transition rate depending on both the temperature
and the interstitial concentration has been calculated. Then, using the obtained
result, we consider the model of surface diffusion that reproduces qualitatively
the diffusion behaviour of the hydrogen isotopes on $W(110)$ surface, observed in the
lowtemperature experiments [S.C. Wang and R. Gomer, J. Chem. Phys. {\bf 83}, 4193 (1985)].
The coverage dependence of the diffusion coefficient is
determined by Monte Carlo simulation. This model allows to suppose that
substantially the different diffusion of hydrogen (tritium) and deuterium can be the
outcome of their different quantum statistic and the direct interactions between
the adparticles.
\end{abstract}

\section{INTRODUCTION}
Quantum diffusion both in volume of solids and on solid surface has been attracting
considerable attention for the past several decades mainly due to extraordinary
high mobility of interstitial hydrogen and its isotopes at low temperatures.
Theoretical consideration of this problem rests on low concentration of interstitial particles.
This allows to reduce duffusion processes to tunneling dynamics of syngle particle,
moving in an effective potential and interacting with the solid environment,
in particular, with phonon sybsystem of the perfect lattice \cite{1,2,3,4,5,7,8,9,10,11,12,13,14,%
15,16,17,18,19,20,21,22,23}.
In different approaches interection with environment is assumed to be either
linear or bilinear with respect to phonon variables and considered within the scope
of perturbation theory. As a result of tunneling a particle transfers from one
interstitial site to an adjacent that. Near $T = 0$ corresponding motion  is
due to band-like propogation (coherent diffusion). At increased temperatures
localization of a particle in a site occurs by interaction with phonons and the
hopping regime is realized (incoherent diffusion). In a metal at very low
temperatures the tunneling dynamics of particles proves to be  strongly
depending on nonadiabatic interection with the conduction electrons \cite{24,25,26}.

To  calculate the diffusion coefficient different approahes have been suggested to a certain
extent using the above physical pictures.  The review of some theoretical methods
and subsequent generalizations are given in the papers \cite{27,28}.

At the finite impurity concentrations the interaction between diffusing particles
can distort the low concentration picture. So, Kagan and Maksimov \cite{7} have shown that
when particle concentration exceeds some low critical value elastic interaction
between particles at $T = 0$ results in suppression of the band motion and in
complete particles localization.

At high interstitial concentrations the diffusion process is to be determined
to a great extent by short-range interactions between diffusing  particles
\cite{8}. An interesting example in this sense is the unusial diffusion of
hydrogen isotopes on the $(110)$ surface of tungsten at low temperatures
revealed by DiFoggio and Gomer \cite{29} and Wang and Gomer \cite{30}. They
have found that dependence of the diffusion coefficient $D$ on coverage $\theta$
in all cases displays oscillator-like  behaviour. However while $D$ for
hydrogen (and tritium) decreases  at $\theta \to 1$, the diffusion coefficient
for deuterium increases. This phenomena has not been explained up to now.
Muttalib and Sethna \cite{5} and Gomer \cite{31} have supposed that it may
be a consequence of the difference in quantum statistics of $^1$H and $^2$H.
From our point of view a double effect takes place  here: the existence of
interaction between diffusing adatoms and difference in statistics.

In this work we first study quantum diffusion in a valume of a crystalline
solids at high concentrations of interstitials, when the hopping regime
is realized. Because of the large distortion of lattice the corresponding
temperature region can extend to lowerest  temperatures. We confine ourselves
to a temperature region in which electronic degrees of freedom can be
eleminated from  consideration within of adiabatic approximation. Allowing
for short-range interaction between interstitials, we find an expression
for the transition rate depending on concentration. Then a model of surface
diffusion , showing the same behaviour as it does for hydrogen and deuterium
on $W(110)$, is considered. The Monte Carlo method has been utilized
to simulate surface diffusion of adatoms and to calculate the coverage
dependence of the tracer diffusion coefficient. In conclusion on the basis
of the considered model we discuss the reasons for difference in diffusion
of $^1$H and $^2$H atoms on $W(110)$.

\section{TRANSITION RATE}
When considering the diffusion process in a crystal at a low interstitial
concentration one proceeds from the supposition that the jumps are
sufficiently rare events, so  in  the intervals between the jumps a
thermodynamic equilibrium is reached and the system vibrates around a stable
crystalline configuration. The same situation occurs at high concentrations.
In this case the mobility of particles will be determined by their jumps over
the vacant interstices, whose concentration is small. So, between the jumps
in such a crystal thermodynamic equilibrium is also reached  and  a system
will vibrate  around the corresponding crystalline configuration.

Let the vibrational states of the system (host atoms plus interstitials) be
described  by the wavefunctions $\psi^{(r)}_{\{n\}}$ , depending on the
configuration  of the occupied interstices $r = ({\bf r}_1, {\bf r}_2, \dots)$,
with $\{n\}$ denoting the set of quantum numbers. $\psi^{(r)}_{\{n\}}$  take no
account of the jumps and are the eigenfunctions of the some Hamiltonian
$H_0 = T + U^{(r)}_0$.

For different configurations $r$ and $r^\prime$,  $\psi^{(r)}_{\{n\}}$ and
$\psi^{(r^\prime)}_{\{n^\prime\}}$ are localized in different regions of the
configuration space and their overlap is small \cite{6}.  Considering these
states as approximately ortogonal we can write down the transition rate as \cite{32}
\begin{eqnarray}
W = && <W^{(r r^\prime)}>^r \, ,\\ \label{1}
W^{(r r^\prime)} = && \bigl<\,\frac{2\pi}{\hbar}\sum_{\{n^\prime\}}\bigl|<\psi^{(r^\prime)}_{\{n^\prime\}}\bigl|H - H_0\bigr|\psi^{(r)}_{\{n\}}>\bigl|^2
\delta(E^{(r^\prime)}_{\{n^\prime\}} - E^{(r)}_{\{n^\}})\,\bigr>_{\{n\}} \ \ . \label{2}
\end{eqnarray}
Here and further $r$ and $r^\prime$ denote the initial and final interstitial configurations
during the jump of an interstitial to the nearest adjacent site, $H = T + U$
is the exact Hamiltonian allowing for the jumps; $<...>_{\{n\}}$, $<...>^r$
are the thermal average  over the initial states $\{n\}$ and the average over
all possible initial configurations $r$, respectivly; $E^{(r)}_{\{n\}}$ is the
energy of the state $\psi^{(r)}_{\{n\}}$.

To calculate $W$ let us introduce the following approximations.\\
(i) Taking into account
that the overlap between  $\psi^{(r)}_{\{n\}}$ and  $\psi^{(r^\prime)}_{\{n^\prime\}}$
is small we can simplify the matrix element in Eq.(\ref{2}), using approximation
\begin{equation}
<\psi^{(r^\prime)}_{\{n^\prime\}}\bigl|H - H_0\bigr|\psi^{(r)}_{\{n\}}> = \Delta\Phi^{(r)}<\psi^{(r^\prime)}_{\{n^\prime\}}\bigm|\psi^{(r)}_{\{n\}}> \,. \label{3}
\end{equation}
where $\Delta\Phi^{(r)} = U - U_0^{(r)}$ is calculated at a point of the overlap
range, corresponding to some "transition" configuration of interstitials $r^*$.
In this configuration the jumping interstitial is situated in the middle
of the way  between the initial and the final sites during the jump. The
positions of other particles  for present  remain not determined. It should be noted
that, actually, the difference $U - U_0^{(r)}$ involves only degrees of
freedom related to the jumping interstitial and its nearest neighbours and is notably
distinguished from zero in some finite region of the crystal.\\
(ii) By occupying some interstice an interstitial interacts with the neighbouring
host atoms. The additional  force repulsion from the nearest interstitials must
result in the sizable localization in sites even of such light particles
as hydrogen. In this case one can approximate the states $\psi^{(r)}_{\{n\}}$
by the eigenfunctions $H_0^{(r)}$, taken in the harmonic approximation
\begin{equation}
H_0^{(r)} = \Phi_0^{(r)} + \frac{1}{2}\sum_k m_k\dot{x}_k^2 +  \frac{1}{2}\sum_{k l}\Phi^{(r)}_{kl}(x_k - x^{(r)}_{k0})(x_l - x^{(r)}_{l0}) \, .
\label{4}
\end{equation}
Here $x_1 , x_2 , \dots x_{3N}$ are the coordinates of all $N$ particles of the system,
$x_{k0}^{(r)}$ are the equilibrium positions of particles for the configuration $r$; $\Phi^{(r)}_{kl}$ are the force constants and
$\Phi_0^{(r)}$ is the potential energy of the system at the equilibrium position;
$m_k = m$ is the host atom mass, if $x_k$ relates to the lattice  coordinates
and $m_k = M$ is the interstitial mass, if $x_k$ relates to the interstitial degrees
of freedom.

It is known \cite{33} that when the volume of the crystal $V \to \infty$,
any additive quantity $V^{-1}A^{(r)}$, due to the spatial homogeneity of the system,
has the determinate limit $A_0$ not depending on $r$, i.e.
\begin{equation}
\lim\limits_{V\to\infty}V^{-1}A^{(r)} =  \lim\limits_{V\to\infty}<V^{-1}A^{(r)}>^r = A_0\, .
\label{5}
\end{equation}
As $\Phi^{(r)}_0$  has the property of "self-averaging"  (\ref{5}), taking a
sufficiently large volume $V$, one will has to any degree of accuracy
\[\Phi_0^{(r)} = V\Phi_0 \, .\]
Further we shall take $\Phi_0 = 0$.

Procceding to the normal-mode coordinates $q^{(r)}_k$, one writes down Hamiltonian
(\ref{4}) as
\begin{equation}
H_0^{(r)} = \frac{1}{2}\sum_s m{\dot q}_s^{(r)2} + \frac{1}{2}\sum_sm{\omega_s^2}q_s^{(r)2}\, .
\label{6}
\end{equation}
Here we have also allowed for the density of the vibrational states (distribution
function of the frequencies of the normal modes) ${\nu}^{(r)}(\omega)$  has the property
(\ref{5}), i.e
\begin{equation}
\lim\limits_{V\to\infty}\nu^{(r)}(\omega) =  \lim\limits_{V\to\infty}<\nu^{(r)}(\omega)>^r = \nu (\omega , c)\, ,
\label{7}
\end{equation}
where $c$ is the interstitial concentration (the number of the  interstitials
 divided by the number of the interstices). Thus, for a large enough crystal the
frequency spectrum does not practically depend on the configuration of the
occupied interstices $r$.

In the paper \cite{32} it has been demonstrated that  the overlap integral between the
eigenfunctions of Hamiltonian (\ref{6}), considered in the configuration space,
one can represent  as
\begin{equation}
<\psi^{(r')}_{\{n'\}}|\psi^{(r)}_{\{n\}}>  =  \prod_k\,\int\,dq_k\varphi_{n_k}(q_k+b^{(rr')}_k)\varphi_{{n'}_k}(q_k),
\label{8}
\end{equation}
where $\varphi_n(q_k)$ are the normalized harmonic oscillator eigenfunctions with
the frequency $\omega_k$ and the mass m and
\[ |b^{(rr')}_k| = a\bigl(M/m\bigr)^{\frac{1}{2}}|\alpha^{(r)}_{sk}|\sim a\bigl(M/m\bigr)^{\frac{1}{2}}\,\alpha(M/m)\bigg/\sqrt{3N}\, .\]
Here  $(M/m)^{\frac{1}{2}}\alpha^{(r)}_{sk}$ is the matrix connecting the displacement
$x_s - x_{s0}^{(r)}$ with the normal coordinates $q_k^{(r)}$; $a$ is the jumping
distance, $\alpha(M/m)$ is some dimensionless function and the interstitial is assumed to hop along
the $x_s$ direction, i.e. $|x_{s0}^{(r)} - x_{s0}^{(r')}| = a$.

The matrix elements in (\ref{8}) are proportianal to the factor $J = e^{-\lambda}$ ,
where $\lambda \sim (a^2/<u_i>_0) \gg 1$ and $<u_i>_0$  is the mean square of the
displacements of the zero vibrations interstitials.

Allowing for approximation (\ref{3})  and inserting (\ref{8}) into (\ref{2})
one can calculate $W^{(rr')}$ employing the small-polaron formalism \cite{34}.
Taking limits $N\to\infty\, , V\to\infty$ and allowing for the fact that $\Delta\Phi^{(r)}$
does not depend on either $N$ or $V$, one obtains \cite{35}
\begin{equation}
W^{(rr')}=\frac{|\Delta \Phi^{(r)}|^2}{\hbar^2} \,e^{-S_T(c)}
\int\limits_{-\infty}^{\infty}{dt[{e^{G(t;c)}-1}]},
\label{9}
\end{equation}
where
\[ S_T(c)=\frac{\alpha^2 a^2 M}{2\hbar}\int{d\omega\, \nu(\omega, c)\,\omega\,
   \coth \frac{\hbar\omega}{2k_B T}},\]
\[ G(t; c)=\frac{\alpha^2 a^2 M}{2\hbar}\int{d\omega\, \nu(\omega, c)\,\omega\,
   \csc\frac{\hbar\omega}{2k_B T}\, \cos\omega t}.\]
Now the averaging over the different configurations $r$ in Eq.(\ref{9}) is to be
carried out in the  finite region where $\Delta\Phi^{(r)}$  is notably distinguished from
zero. To carry out the averaging, one  notes that in the first approximation
$\Delta\Phi^{(r)}$ is an anharmonic  contribution in  the potential energy of an
interaction between the jumping particle, localized in the middle of the way
between the initial and the final sites during the jump, and its nearest environment.
Let $n_k^{(r)}$ be the number of $k$-th nearest interstitials of this particles, 
corresponding configuration $r$. Then one can write down
\begin{equation}
\Delta\Phi^{(r)} =  \Delta\Phi + \sum_k n_k^{(r)} V^{(2)}_k \, ,
\label{10}
\end{equation}
where  $\Delta\Phi$ is the contribution of the host atoms to  $\Delta\Phi^{(r)}$
and $V^{(2)}_k$ is the contribution of two-body interactions between the jumping
particle and one of its $k$-th neighbours among of the intertstitials.

The Eq(\ref{10})  shows that $\Delta\Phi^{(r)}$ depends only on occupation numbers
$n_k^{(r)}$ rather than on placing of interstitials over equivalent interstices.
It is easy to see that the averaging $|\Delta\Phi^{(r)}|^2$  over the different
configurations is equivalent to averaging of the numbers $n_k^{(r)}$ over the
binomial distribution with the probability $p = c$ that the giving interstice
is occupied and the probability of the contrary event $q = 1 - c$, independly for
every $k$. The random variables $n_k^{(r)}$ take  the following values
$n_k^{(r)} = 0, 1, 2\dots \overline{n}_k$, where $\overline{n}_k$ is the maximum possible
number of   $k$-th neighbours among of the interstitials.

In addition let there exist also  three-body interactions between  interstitials,
for example, with the participation of the first two neighbours of the jumping
particle. If $V_1^{(3)}$ is the  average contribution to $\Delta\Phi^{(r)}$ from
each one of such three-body configurations the one can write
\begin{equation}
\Delta\Phi^{(r)} =  \Delta\Phi + \sum_k n_k^{(r)} V^{(2)}_k  +
\frac{1}{2}n_1^{(r)}(n_1^{(r)} - 1)V_1^{(3)} + \dots \, ,
\label{11}
\end{equation}
where dots denote the contributionss from thre-body interactions with the particuipation
of the following neighbours and also the possible contributions from $n$-body interactions
with $n > 3$.

The Eq.(\ref{11}) depends on the occupation numbers only and its square
has the following form
\begin{equation}
|\Delta\Phi ^{(r)}|^2=|\Delta\Phi |^2 \left (1 +\sum_i b_i n^{(r)}_i + \sum_{ij} b_{ij}\,
n^{(r)}_i n^{(r)}_j + \sum_{ijk} b_{ijk}\, n^{(r)}_i n^{(r)}_j n^{(r)}_k +... \right ),
\label{12}
\end{equation}
where $b_i, b_{ij}, b_{ijk},\dots $ are the known dimensionless coefficients.

The Eq.(\ref{12}) can be averaged over the configurations as it has been indicated
above. Taking into account that the average $<n^l>$, where $n$ are the occupation
numbers of the interstices, over the binomial distribution is the polinomial in $p = c$
of degree $l$, one obtains the transition rate $W = <W^{(rr')}>^r$
\begin{equation}
W(T,c)=Q_{2n-2}(c)W_0(T,c)\, ,
\label{13}
\end{equation}
where
\[ W_0(T,c)=\frac{|\Delta\Phi|^2}{\hbar^2}\, e^{-S_T(c)}
 \int\limits_{-\infty}^{\infty}{dt[{e^{G(t;c)}-1}]}\, ,\]

\begin{equation}
Q_{2n-2}(c)=1+ a_1c + a_2c^2 +...+ a_{2n-2}c^{2n-2}\, \, .
\label{14}
\end{equation}
Index $n$ in Eq.(\ref{14}) indicates that the interstitial's potential energy
contains $n$-body potentials as well.

\section{MODEL OF QUANTUM DIFFUSION ON SOLID SURFACE}
The expression for transition rate obtained in the preceding section can be applyed to
the study of the diffusion on solid surface without any restrictions. In this
section we shall consider a model system , in which depending on parameters
of the model, surface diffusion of the particles exhibits the same behaviour
as the diffusion of hydrogen and deiterium on W(110) \cite{29,30}.
First we calculate the frequency spectrum of the defective crystal to find $W$ from Eq.(\ref{13}),
employing Dean's direct numerical method \cite{36,37}. Then the coverage dependence of
the tracer diffusion coefficient is determined by Monte Carlo simulations
on a fixed two-dimensional lattice.

\subsection{Frequency spectrum}
We use a model in which  the solid surface is represented by the cubic lattice of
size $n\times l\times 3$, consisting  of three layers, with regular sites
occupied by substrate atoms of mass $m$. The adsorbed particles of mass $M$
and with desired coverage $\theta$ are localized in the interstices of the
crystal lattice ranging in the upper layer. In our model the particles are connected to their nearest neighbours by central and
noncentral harmonic springs. This choice affords the simplification that the motions in each
cartesian direction are independent \cite{36,37}.

In this section we denote the displacement of the $i$-th particle in $x$-direction as
$x_i$. The enumeration of the particles as well as the force constants connecting
the nearest neighbours  are shown in Figs.\ \ref{fig1} and \ref{fig2}, where
the notation is introduced
\begin{eqnarray}
\left[s\right]&=&n_1 + n_2 +\dots + n_{s-1} + 2(s - 1)n\, ,\hspace{1cm} 1 < s < l + 1 , \nonumber  \\
\left[1\right] &=& 0\, .  \nonumber
\end{eqnarray}
The difference $p_s = n_s - n$ is the number of the interstitials localized in
the interstices of the $s$-th vertical column (along  $y$-direction).

Let $k_{ij}$ be the harmonic force constant connecting  the nearest neghbours $i$ and $j$.
Then  the equation  of motion for $i$-th particle can be written as
\begin{equation}
m_i \ddot{x}_i = {\sum_j}^{\prime} k_{ij}(x_j - x_i) ,\hspace{1cm} k_{ij} =  k_{ji}\, .
\label{15}
\end{equation}
The prime at the sum denotes the omission of the term with $j = i$ . In addition the periodic
boundary conditions were imposed  in $y$-direction and rigid wall boundary conditions in the $x$-direction.
For simplicity it has been assumed that the adparticles and
substrate atoms of the lower layer do not interact. Thus  there exist the
lateral interactions for adparticles only.

Allowing for ${\ddot x}_i = - \omega^2 x_i$ and introducing the notations
$u_i = m_i^{1/2}x_i$, ${\overline\omega}^2 = \omega^2/\omega^2_0$, $\omega^2_0 = 4k/m$,
one has instead of (\ref{15})
\begin{equation}
{\overline\omega}^2 u_i = \alpha_iu_i +  {\sum_j}^{\prime} \beta_{ij}u_j \, ,
\label{16}
\end{equation}
where
\[\alpha_i = \frac{m}{m _i}{\sum_j}^{\prime} \frac{k_{ij}}{k};\hspace{1.5cm}\beta_{ij} = -
\frac{m}{4k}\frac{k_{ij}}{\sqrt{m_im_j}} \, . \]
The system of the equations (\ref{16}) is equalent to the eigenvalue problem
\begin{equation}
{\bf M}^{(r)}\cdot{\bf u} = {\overline\omega}^2 {\bf u}
\label{17}
\end{equation}
for the symmetric matrix ${\bf M}^{(r)}$ of order $N\times N$, depending on the configuration of the
occupied interstices and having a block tridiagonal form
\begin{equation}
 {\bf M}^{(r)} = \left( \begin{array}{ccccccc}
   {\bf A}_1   &{\bf B}_2   &        &           &      &          & 0     \\
   {\bf B}^T_2 &{\bf A}_2   &{\bf B}_3 &           &      &          &       \\
             &{\bf B}^T_3 &{\bf A}_3 & {\bf B}_4   &      &          &       \\
             &          &   .    &    .      &  .   &          &       \\
             &          &        &    .      &  .   &    .     &       \\
             &          &        &           &  .   &    .     &   .    \\
     0       &          &        &           &      &{\bf B}^T_l &{\bf A}_l  \
\end{array}\right) \, .
\label{18}
\end{equation}
Here ${\bf B}^T_s$ is the transpose of ${\bf B}_s$. The explicit forms  of submatrices
${\bf A}_s$ and ${\bf B}_s$ are given in the Appendix.

To calculate the spectrum of eigenvalues of the matrix ${\bf M}^{(r)}$ we use Dean's
direct numerical  method. Originally it was applied for finding the frequency
spectrum of the disorded lattices with substitution impurities and has
been discussed in detail in Refs.\cite{36,37}. Here  only the main poins
are outlined.

First one finds the integrated frequency spectrum $N({\overline\omega}^2)$
which gives the number of the squared frequencies less than some ${\overline\omega}^2$.
If $\eta({\bf X})$ denotes the number of negative eigenvalues of a matrix ${\bf X}$,
then $N({\overline\omega}^2) = \eta({\bf M}^{(r)} - {\overline\omega}^2{\bf I})$
and ${\bf I}$ is the unit matrix.  Calculation of $\eta({\bf X})$ , where ${\bf X}$
is the matrix of form (\ref{18})  is based on the negative eigenvalue theorem by Dean and
Martin \cite{36}.  In the next step the squared frequency distribution function
$G({\overline\omega}^2)$ is determinated by
\[
G(\bar\omega^2)=\frac{N(\bar\omega^2+d\bar\omega^2)-N(\bar\omega^2)}
{d\bar\omega^2N}
\]
After $G(\bar\omega^2)$ has been found the frequency spectrum  $\nu(\omega)=2\omega G(\bar\omega^2)/{\omega^2_0}$.
is calculated. The system's sizes are picked so that  $\nu(\omega)$ depends only on a coverage
${\theta}$ rather than the configuration of the occupied interstices.

To calculate the frequency spectrum we consider the lattice containing $30\times 30\times 3$
regular sites. The adparticles are distributed at random on interstices with the given
probability $\theta$. At the picked lattice's size the frequency spectrum practically
does not depend on particular adparticle configuration. The following ratios
between the force constants have been taken: $k_1/k = k_2/k = 1$, $k_3/k = 8$, $k_4/k = 9$.
The mass ratio for a substrate atom and adatom has been taken the same as for system
$^1$H$/W$, $^2$H$/W$.  The calculations have been carried out at different
coverages from the region $0.4\leq \theta \leq0.9$ and showed the insignificant variation of the
distribution function $G({\overline\omega}^2)$  at the increase of $\theta$ in
both cases. The typical histograms for $G({\overline\omega}^2)$  are showed in Figs.\ \ref{fig3} and \ref{fig4}.

\subsection{Monte Carlo simulation of surface diffusion}
For simulation of the diffusion process on the surface we consider lattice
gas of the $N$ adsorbed particles occupying at the regular sites of $L\times L$
square lattice. In our model Hamiltonian of the system has been taken in the
form allowing for two and three-body interactions between adatoms
\begin{equation}
 H=J_1\sum_{(NN)}c_ic_j+J_2\sum_{(NNN)}c_ic_j+J_t\sum_{(t)}c_ic_jc_k\, .
\label{19}
\end{equation}
Here $J_1>0$ and $J_2<0$ are the nearest $(NN)$  and next - nearest neighbour
$(NNN)$ interaction energies, $J_t>0$ is three - body interaction energy (t).
Corresponding adatom configurations are shown in Fig.\ \ref{fig5}. The positive
constant $J_n > 0$ corresponds to repulsion. $c_i$ is the occupation number
of site  $i$  and takes the value $1$ and $0$ for full and empty site
respectively. Double occupancy of sites is excluded.

Within the scope of the random walk theory the tracer diffusion coefficient
for the twodimensional system is defined as
\begin{equation}
D^*=\lim_{t\to\infty}\frac{1}{4Nt}\sum^N_{i=1}<\Delta r^2_i(t)>\, ,
\label{20}
\end{equation}
where $t$ is the elapsed  time and $<\Delta r^2_i(t)>$ is the mean-square
displacement of the $i$-th particle.

To calculate $D^*$ Monte Carlo simulation of surface  diffusion has been
carried out in the canonical ensemble applying  the Metropolis importance
sampling  \cite{38}. The actual procedure was the following. An initial
adatom configuration corresponding to the desired adatom coverage $\theta$ was
given by the random pick of $\theta L^2$ sites. Then the temperature $T$
was established and employing the standard  procedure a large  number of Monte
Carlo steps (MCS) were performed to reach the thermodynamic equilibrium.
The equilibrium was assumed to be reached when the total energy of the system
started to fluctuate about its average value. Typically in a lowtemperature
region about $10^3$ MCS's are  required to establish equilibrium in the
lattice containing $50\times 50$ sites at $0.4 <\theta < 0.9$ for  $J_ 1 = 2$ kcal/mol,
$J_2 = - 1.5$ kcal/mol, $J_t = 1$ kcal/mol .

After thermodynamic equilibrium had been established a diffusion run was started.
Here we follow the algoritm employed in Refs. \cite{39,40}. The elementary step of surface
diffusion is assumed to be a jump of an adsorbed particle from the occupied initial site
to an empty nearest neighbour one. First of all, the initial site of lattice is picked
at random, if filled, an adjacent final site is randomly choosen. If this site
is empty the jump can occur  with some probability $P_J$, otherwise no jump
occurs and a new initial site is picked. One MCS corresponds to $L^2$ the
random picks of the initial site for the jump.

The time scale is defined by the span $\Delta t$ in which an adatom is allowed
to attempt a jump to an adjacent site. If the mean adatom's lifetime in a site
is $\tau = 1/W$ , where $W$ is given by Eq.(\ref{13}), the probability of
a jump during $\Delta t$ one can write as
\[P_j = \frac{\Delta t}{\tau}\, .\]
In order to reduce computational time we have taken $\Delta t = 1/W_{max}$  with
$W_{max}$ = max$W(T,\, \theta )$ in the region of values $T$ and $\theta$ we are
interested in. Thus the jump probability is given by
\begin{equation}
P_J=\frac{W(T,\, \theta)}{W_{max}} \, .
\label{21}
\end{equation}
The choice of (\ref{21}) means that when $W = W_{max}$  the jump to the
nearest vacant site occurs with the unity probability.

To calculate  $W = Q_{2n-2}W_0$ we have used the frequency spectrum
obtained in the preceding section and have taken $n =3$ that corresponds to the existence
of pairwise and triplet interactions of adatoms. To reproduce the behaviour
of hydrogen isotopes in the diffusion experiments we consider two model expressions for
$Q_4$
\begin{equation}
Q^{(1)}_4(\theta)=1-0.6\,\theta+6.1\,\theta^2-11.8\,\theta^3+5.7\,\theta^4
\label{22}
\end{equation}
that corresponds to $^1$H and
\begin{equation}
Q^{(2)}_4(\theta)=1-9.9\,\theta+62.2\,\theta^2-122.1\,\theta^3+73.6\,\theta^4
\label{23}
\end{equation}
for $^2$H. Their behaviour is shown on Fig.\ \ref{fig6}. The shape of the curves
are simular but while the minimum of $Q^{(2)}_4$ is located in the range
$\theta < 1$  the minimum of $Q^{(1)}_4$ is displaced into the range $\theta > 1$ .
The coefficients of the polynomials are chosen in such way that the functions $Q_4^{(i)}$
should have the given values at the points of extremum. The remaining indeterminacy
in the coefficiets practically does not affect the behaviour of $D^*$ .

To calculate the diffusion coefficient $D^*$ diffusion runs of 1000 MCS's
for $40$ different initial adatom configurations were performed. The results
of the calculation of coverage  dependence of $D^*$  at $40$K and $\alpha = 1$ (see Eq.(\ref{9}))
are presented on Fig.\ \ref{fig7}. Figs.\ \ref{fig8} and \ref{fig9} show the
temperature dependence of $D^*$ for the different coverages.

\section{DISCUSSION}
In this work we have considered quantum diffusion  in a crystalline solid at high
interstitial concentrations. Our main result is the equation (\ref{13}) for the
transition rate taking into account both the temperature and the concentration dependences.
Futher we have studied  the model system  imitating quantum diffusion of
adatoms on the solid surface. The diffusion behaviour  found in this model can be used
for qualitative explanation of the experimental situation revealed in diffusion of
hydrogen  isotopes on $W(110)$ surface \cite{29,30}.

It is known that the value  measured in an experiment is the chemical diffusion
coefficient $D$, given by Kubo-Green formula. For qualitative discussion we can
make use the approximate expression \cite{31}
\[D\sim \frac{<N>}{<(\delta N)^2>}D^*\, ,\]
where $<(\delta N)^2>/<N>$ is the normalized mean-square fluctuations of the
adatom's number. In the region of low temperatures and high coverages the
thermodynamic factor   $<N>/<(\delta N)^2>$  is insignificantly altered,
so the behaviour of $D$ is mainly determined by the tracer diffusion coefficient
$D^*$  whose coverage dependence is shown on Fig.\ \ref{fig7}.  Qualitatively the curves
presented  on this figure show the same behaviour as the experimental  curves
 for $^1$H and $^2$H from Ref.\cite{30}.

The coverage dependence of $D^*$ is
mainly determined by the polynomial $Q_4(\theta)$. The choice of $Q_4 = Q_4^{(1)}$
corresponds to\, $^1$H diffusion and $Q_4 = Q_4^{(2)}$ to\,  $^2$H diffusion. As
the coefficients of $Q_4 = <|\Delta \Phi^{(r)}|^2>^r$ depend on "transition" configuration $r^*$,
the difference between  $Q_4 = Q_4^{(1)}$ and $Q_4 = Q_4^{(2)}$ indicates that these
configurations for $^1$H and  $^2$H are to be different. Indeed, the configuration $r^*$
for the present  is not exactly defined. The additional condition can consist in giving
of the distance $R$ between the neighbouring  adatoms involved in the diffusion jump. It is
natural to take  $R$ equal to the average distance between  the adatoms in the
states discribing particle's transition to the neighbouring site. The corresponding
wavefunctions are to allow for a symmetry in the permutations of particles.
Owing to the difference in the statistics  the close approaches
are more likely for bosons (deiterium) than  for fermions (hydrogen and tritium).
So during the jump the average distance between $^1$H atoms turns out more
 than for $^2$H. As a result the the configurations $r^*$  shall be different
in these two cases.

Further, the coefficients  of $Q_4$  are determined
by the potentials of interactions  between the diffusing particles. When
the distance between particles is small  the interaction's potentials  can
take  very large value. Then  it follows from the above that the
coefficients of $Q_4^{(2)}$  for deiterium are to be more (in magnitude) than the
coefficients of $Q_4^{(1)}$ for hydrogen. As a result polynomial  $Q_4^{(2)}$
is rapidly altered and  starts to rise at $\theta\to 1$  as it is shown
on Fig.\ref{fig6}. The diffusion coefficient for deiterium demonstrates the
same behaviour.

Fynally we can imagine diffusion motion of $^1$H and $^2$H as follows. Approaching
closely $^2$H atoms suffer the pronounced repulsions that results in the
acceleration of the diffusion process. The decisive contribution to the
observed diffusion behaviour at the high coverages comes from three (or more)-body
interactions, whose potential , is probably, rapidly vanishing  with increasing
distances between particles. So mainly $^2$H atoms acquire the pronounced
energies in the three-body interactions. It is the existence of three-body
interactions that dominates the rise of $D$ for $^2$H  at $\theta\to 1$.
At the diminished coverages the diffusion motion is mainly dominated by the two-body
interaction between particles and results in the same behaviour of $D$ for
$^1$H and $^2$H  as  it has been revealed in the experiment.

\appendix
\section*{}
The submatrices ${\bf A}_s$ and ${\bf B}_s$ from Eq.(\ref{18}) have the block form
\begin{equation}
{\bf A}_s = \left(
\begin{array}{ccc}
{\bf A}_s'  &{\bf C}_s   &     0  \\
{\bf C}^T_s &{\bf G}_1   &\bf  C  \\
     0    &{\bf C}^T   &{\bf  G}_2
\end{array}\right),
\label{a1}
\end{equation}
\begin{equation}
{\bf B_s} = \left(
\begin{array}{ccc}
{\bf B}_s' &      &    0 \\
         &\bf B &      \\
 0       &      &\bf B
\end{array}\right).
\label{a2}
\end{equation}
The form of the submatrices from (\ref{a1}) and (\ref{a2}) strongly depends on the
configuration of the occupied interstices, so here we shall show their
structure only.

${\bf A}_s$ is a square matrix of five diagonal banded form with two additional elements
in the upper right and lower left  corners
\[
{\bf A}_s' = \left(
\begin{array}{cccccc}
\alpha_{[s]+1} & \beta_{[s]+2} & \gamma_{[s]+3}   &                &                                 & \beta \\
 .              &      .        &       .           &    .         &                              &    \\
 .              &      .         &       .        &      .         &               .                 &    \\
               &      .         &       .         &     .         &              .                &   .  \\
	       &                &          .       &      .        &              .                &  .   \\
\beta          &               & \phantom{\gamma_{[s]+3}}         & \gamma_{[s]+n_s}&\beta_{[s]+n_s}&\alpha_{[s]+n_s} \\
\end{array}\right),
\]
where $\beta =-{k_1}/{4k} \,\,\, \beta_i=\beta_{i-1,i},\,\, \gamma_i=\beta_{i-2,i} $.

${\bf C}_s$ is a matrix of order $n_s\times n$. Its nonzero elements are on
$p_s$ diagonals
\[
{\bf C}_s = \left(
\begin{array}{ccccccc}
c^{(s)}_{11}   &\phantom{c_{11}} &\phantom{c_{11}}   &\phantom{c_{11}}  &\phantom{c_{11}} & \phantom {c_{11}}  &        0       \\
           .   &   &    &   &  &   &                \\
           .   & .  & . &   &  &   &                \\
           .   & .  & . & . &  &   &                \\
c^{(s)}_{p_s1} & .  & . & . &. &   &               \\
               & . & .  & . &. & .&                \\
               &   & .  & . & .& .   &c^{(s)}_{nn} \\
               &   &    & . & .&  .  &          .      \\
               &   &    &   & .& .   &          .       \\
               &   &    &   &  & .  &         .       \\
        0      &   &    &   &  &   &c^{(s)}_{n_sn}
\end{array}\right),
\]
where $c^{(s)}_{ij}=\beta _{[s]+i,[s]+n_s+j}$.

The matrix $G_i$ $(i = 1,2)$   is of tridiagonal form, of order $n\times n$
with additional elements in the upper right and lower left corners.
\[
{\bf G}_i = \left(
\begin{array}{ccccccc}
\bar\alpha_i & \beta        &      &      &   &       & \beta      \\
\beta        & \bar\alpha_i &\beta &      &   &       &           \\
            &       .       &  .   &   .  &   &       &          \\
            &              &  .    &   .  & . &       &       \\
            &              &       &   .  & . &   .    &       \\
            &              &       &      & . &   .    &\beta       \\
\beta       &              &       &      &   & \beta &\bar\alpha_i\\
\end{array}\right).
\]
If the index $s$ does not relate to the side i.e. $(s\neq 1,l)$ , then
\[\bar\alpha_1=\frac{1}{2k}(k_1+k_2+k),\hspace{1.5cm}
  \bar\alpha_2=\frac{1}{2k}(k_1+\frac{k_2}{2}+k).\]
On sides $\bar\alpha_i$ are to be altered insignificantly.

The matrix ${\bf C} = -(k_2/4k){\bf I}$ is of order $n\times n$ .  The matrix ${\bf B}_s'$ is
of order $n_s\times n_{s+1}$. Its  nonzero elements are on $p_s + p_{s+1}$
diagonals.
\[
{\bf B}_s' = \left(
\begin{array}{cccccccccc}
\beta^{(s)}_{11}   & . & .& . &\beta^{(s)}_{1p_{s+1}} &  &\phantom{\beta_{11}} &\phantom{\beta_{11}} &\phantom{\beta_{11}}  &  \\
 .   &.  & . & .  &  . & .  &   &  & &  \\
 . & . & . &.  & . & .&.  &  & &   \\
.  & .  & . & .  &. & . &  . &.  & & \\
\beta^{(s)}_{p_s1} & .& . &  . & .  & . & .  & .  &. &  \\
                   & . & . & .  &   .  &  . & . & .  & . &\beta^{(s)}_{nn_{s+1}} \\
                   &   &.  & .  &  . &. & . & . &.  &. \\
                   &   &  & .  &  . &   .  & . &   & . &   . \\
                   &   &  &    &  . &   .    & .  & .  & . &    .          \\
                   &\phantom{\beta_{11}}  &\phantom{\beta_{11}}  & \phantom{\beta_{11}}  &  &\beta^{(s)}_{n_sn}  &.  &.  & . &\beta^{(s)}_{n_sn_{s+1}}
\end{array}\right),
\]
where $\beta^{(s)}_{ij   }=\beta_{[s]+i,[s+1]+j}$.

Finally  ${\bf B} = (1   /4){\bf I}$  is the matrix of order $n\times n$.

%
%
\begin{figure}
\caption{Enumeration of the particles on the crystalline surface.The harmonic force
constants connecting the nearest neighbours are showed by arrows.}
\label{fig1}
\end{figure}

\begin{figure}
\caption{Enumeration of the particles in the layers perpendicular to the surface.
$k_2$ is the force constant connecting the substrate atoms of upper and lover layers.}
\label{fig2}
\end{figure}

\begin{figure}
\caption{Squared frequency spectrum for $30\times 30\times 3$ lattice at
$\theta = 0.7$, corresponding to $^1$H/W system.}
\label{fig3}
\end{figure}

\begin{figure}
\caption{Squared frequency spectrum for $30\times 30\times 3$ lattice at
$\theta = 0.7$, corresponding to $^2$H/W system.}
\label{fig4}
\end{figure}

\begin{figure}
\caption{Schematic representation of the interaction constants.}
\label{fig5}
\end{figure}

\begin{figure}
\caption{Coverage dependences of $Q_4^{(1)}$ and $Q_4^{(2)}$. }
\label{fig6}
\end{figure}

\begin{figure}
\caption{Coverage dependence of the tracer diffusion coefficient $D^*$ at $40$K for
$J_1 = 2$ kcal/mol, $J_2 = - 1,5$ kcal/mol, $J_t = 1$ kcal/mol;
$\Box $ corresponds to $^1$H/W system, O to $^2$H/W system.}
\label{fig7}
\end{figure}

\begin{figure}
\caption{The temperature dependence of the tracer diffusion coefficient $D^*$,
corresponding to $^1$H/W system, at the different coverages.}
\label{fig8}
\end{figure}

\begin{figure}
\caption{The temperature dependence of the tracer diffusion coefficient $D^*$,
corresponding to $^2$H/W system, at the different coverages.}                  \label{fig9}
\end{figure}

%
%

\end{document}